\documentclass{elsart}
\usepackage{amsmath}
\usepackage{amssymb}
\usepackage{graphicx}

\begin{document}
\begin{frontmatter}

\title{Characterizing the Shapes of
Galaxy Clusters Using Moments of the Gravitational Lensing Shear}

\author{A.E. Schulz$^{1,2}$, Joseph Hennawi$^2$, Martin White$^{2,3}$}

\address{$^1$Harvard University}
\address{$^2$University of California, Berkeley}
\address{$^3$Lawrence Berkeley Laboratory}

\begin{abstract}
We explore the use of the tangential component of weak lensing shear to
characterize the ellipticity of  clusters of galaxies.  We introduce an
ellipticity estimator, and quantify its properties for isolated clusters from
$\Lambda$CDM N-body simulations.  We compare the N-body results to
results from smooth analytic models. 
The expected distribution of the estimator for mock observations is presented,
and we show how this distribution is impacted by contaminants such as noise,
line of sight projections, and misalignment of the central galaxy used to
determine the orientation of the triaxial halo.  We examine the radial profile
of the estimator and discuss tradeoffs in the observational strategy to
determine cluster shape. 
\end{abstract}


\end{frontmatter}

\section{Introduction}

Clusters of galaxies have long been studied in a variety of contexts as
probes of the large scale structure in the universe.
It is commonly assumed that clusters are (approximately) isothermal
spheres in hydrostatic equilibrium, but this is clearly only an idealization.
Key to progress in determining accurate masses and testing our models of
structure formation and of gravity is an ability to relax these assumptions
in a manner guided by observations.  In particular it is of some interest
to determine the shapes of the dark matter halos hosting clusters of galaxies;
information that is in principle obtainable from studies of gravitational
lensing.

Cosmological N-body simulations predict that the dark matter in galaxy
clusters collapses into prolate triaxial ellipsoids
\cite{Dubinski:1991bm,Warren:1992tr}.  The conventional belief is that
since the baryons are a small fraction of the total cluster mass,
baryonic physics has a negligible effect on the halo shape out near the
virial radius.  However one group \cite{Kazantzidis:2004vu} has presented
evidence to the contrary from numerical simulations of cluster formation
which include gas cooling and feedback processes.
They suggest that particle orbits are circularized by gravitational
interaction with the deeper potential wells from baryonic cooling in the
cluster's core, leading to an observable change in halo shape even at large
radius.
If borne out, this result would have strong implications for our ability to
accurately model structure formation on Mpc scales.
Such a claim could be tested with data on halo shapes from weak lensing.

Several research groups have looked at characterizing the shape of a
dark matter halo using the weak lensing shear distortion 
\cite{Wilson:1996,Hoekstra:1997dt,Brainerd:2000fd,Natarajan:2000im,Hoekstra:2003pn,Mandelbaum:2005nf}.  
On galactic scales, \cite{Hoekstra:2003pn} reported an observation of halo
flattening.  They find evidence that the dark matter halos are rounder than
the luminous material but that they align well with the light distribution.
Ref.~\cite{Mandelbaum:2005nf} present results from the SDSS which are
consistent with no flattening.
Little observational work exists yet on the shapes of more massive halos,
which will be our focus.

Employing the weak lensing shear to characterize cluster ellipticity is not
without difficulties.  The signal strength in the quadrupole moment is roughly
10 times weaker than the mean tangential shear.  This reduction in signal
to noise implies that to make a significant detection, observations of 
several clusters will have to be averaged.
The quadrupole moment is also more susceptible to observational systematics
(such as the Point Spread Function or PSF anisotropy correction) than is the mean shear.  The 
presence of substructure in the cluster impacts the measurement substantially, 
as does the projection of other objects in the light cone.
On the other hand by working on the high-mass end of the mass function we
are maximizing the signal from any given object, and by working with clusters
we do not need to worry about the number of objects hosted by any given
dark matter halo: our halo occupation distribution is unity above our mass
threshold.

The goal of this paper is to identify an assayable estimator from a weak
lensing shear observation, $Q$, that will use the azimuthal variation of the
tangential shear to quantify the intrinsic shape of galaxy clusters.
Using N-body simulations of dark matter only, we predict the expected
distribution of $Q$ for 900 mock observations of galaxy clusters,
and we compare the mean result to $Q$ for an elliptical NFW profile.
We quantify the impact of contaminants on the quality of the observable
signal and identify the minimum requirements needed to resolve the recent
controversy on cluster shape.  

\section{Method and analysis}

\subsection{Simulations and maps}\label{sec:sims}

We have used a sample of 30 simulated
N-body clusters.
Each cluster generates 31 weak lensing convergence maps 
obtained by projecting through the volume from different observation 
angles.  The redshift of all clusters is $z=0.411$, the redshift of 
the source plane is assumed to be $z=1$, and both are assumed known.
The simulations and ray tracing are the same as those used in 
\cite{Dalal:2004ut}, while the ray tracing technique is described in 
\cite{Dalal:2003kw}.  The N-body simulations are extracted from a larger
simulation generated with a TPM code \cite{Bode:2003ct}, with a fiducial
cosmology of $\Omega_m=0.3$, $\Omega_{\Lambda}=0.7$, $h=0.7$, $n_s=1$,
and $\sigma_8=0.95$. 
The simulation volume is a cube with sides of length $320
\hspace{0.1 cm}h^{-1}\,{\rm Mpc}$, with periodic boundary conditions.  
The $1024^3$ dark matter particles each have mass $m_p=2.54 \times 10^9 
\hspace{0.1 cm} h^{-1}\,M_{\odot}$.  The cubic spline softening length 
is $\epsilon=3.2 \hspace{0.1 cm}h^{-1}\,{\rm kpc}$.  Clusters are located
using a FoF algorithm with linking length $b=0.2$ in units of the mean
inter-particle spacing.
Clusters are extracted from the simulation volume in sub-volumes of
$5\hspace{0.1 cm}h^{-1}\,{\rm Mpc}$ \cite{Hennawi:2005bm}.

Most of the analysis is carried out on the isolated clusters, but in 
the instances where we investigate the impact of projection effects from 
other objects in the light cone, we have added randomly selected 
cutouts of larger convergence maps calculated with
the N-body simulations described in \cite{Schulz:2002wk} 
to the convergence maps described above.  The fiducial cosmology for 
these simulations is $\Omega_m=0.3$,
$\Omega_{\Lambda}=0.7$, $\Omega_bh^2=0.02$, $h=0.7$, $n_s=1$, and $\sigma_8=1$.

The shear maps are generated from the convergence maps using the 
Kaiser-Squires method, which is non-local and makes use of the Born
approximation \cite{Squires:1995gs}.
In analyzing our simulations we use the reduced shear because it is
the observable quantity \cite{Schneider:1997ge,White:2005jr} and can
differ from the plain shear in regions where $\kappa$ is not infinitesimal.
Since some of our maps contain convergence values of $\kappa>1$  at small 
radii, we have been careful to avoid those regions in the weak lensing shear 
analysis.  We have neglected the rotation effect, but caution that in
regions where $\kappa\sim 1$, this may become important.
The components of the shear are computed from the Fourier transform 
of the convergence $\kappa$ via
\begin{eqnarray}
\gamma_1=\frac{1}{1-\kappa} 
\rm{FT}^{-1} \left[ \frac{k_1^2-k_2^2}{k_1^2+k_2^2}\tilde{\kappa} \right]
\hspace{.1cm}, \hspace{0.5cm}
\gamma_2=\frac{1}{1-\kappa}
\rm{FT}^{-1} \left[ \frac{2k_1k_2}{k_1^2+k_2^2}\tilde{\kappa} \right]
\hspace{.1cm}.
\end{eqnarray} 
where $k_i$ are the components of the Fourier vector conjugate to 2D
sky position and a tilde indicates Fourier transform.

We postulate a family of observable estimators that will quantify the
azimuthal variation of the shear distortion due to the intrinsic shape
of a cluster of galaxies.
We will call the estimator $Q$, and generically, it can be computed from
a shear map by performing the following integral.
\begin{equation}
  Q=\int_0^{2\pi}\int_{\rm annulus} W(\phi)\cdot\gamma_T(r,\phi)
  \hspace{.3 cm} dr d\phi
\label{eqn:Qdef}
\end{equation}
Here $\gamma_T(r,\phi)$ is the tangential component of the weak lensing
shear, and is related to $\gamma_{1}$ and $\gamma_{2}$ by
$\gamma_T(r,\phi)=-{\rm Re}\left[(\gamma_1 +i \hspace{.1cm} \gamma_2)
e^{-i2\phi}\right]$.
In the integral, the angle $\phi$ is defined to be the angle measured from
the semi-major axis of the cluster.  It is assumed that the tangential
shear will be observed in some annulus -- the results from different annuli
can be combined at a later stage if desired.
The overall magnitude of $Q$ scales with $\left\langle \gamma_T\right\rangle$,
which dies off with radius.  To remove this gross trend, we will always 
present $Q$ normalized to $\Sigma_{SIS}$ \cite{peacock-99},
the convergence of a singular isothermal sphere with a velocity 
dispersion of $1000$ km/s.
This normalization is purely for presentation purposes and being known
analytically in advance, keeps the noise properties of $Q$ simpler than
if we had divided by $\left\langle \gamma_T\right\rangle$, which would
have removed the trend more exactly.
Though all radii contain information about the shape, the signal to
noise properties will be better in some observational annuli than in others.
The weight function $W(\phi)$ can in principle be optimized to select
the expected angular variation in tangential shear.  In the case where this
variation is quadrupolar, $W(\phi)=\cos(2\phi)$.  Using $W(\phi)=\cos(2\phi)$
is similar in spirit to the approach presented in 
\cite{Natarajan:2000im,Mandelbaum:2005nf}, 
but without 
the complication of measuring B-mode signal to control PSF systematics.

Using moments of the gravitational lensing shear to characterize the
shape as well as the mass of the underlying dark matter distribution
requires that the principle axes of the 2D projection be known.  For 
realistic levels of noise in the shear measurement, we find that it is 
not possible to determine this direction from the data. 
Fortunately, there exist other indications of the alignment of an
elliptical distribution of dark matter, most notably, the orientation
of the cluster's central galaxy (the brightest cluster galaxy or BCG)
\cite{Dubinski:1997tm}.  
However, in the analysis of
dark matter simulations, we require a method to determine the
the orientation of the semi-major axis, as a reference for measuring 
the angle $\phi$ in the integral above.  The canonical
way of finding the orientation is by computing the moment of inertia
tensor $I_{ab}$ of the dark matter particles in the cluster 
(e.g. \cite{Bailin:2004wu}).
The eigenvectors
of this tensor are oriented along the semi-major and semi-minor axes
of the object, and the square root of the eigenvalues characterize
the cluster's extent along each axis.
In this paper we adopt a similar technique, but compute
the moment of inertia and the resulting
effective ellipticity directly on the 2-D projected $\kappa$ maps.

The simplest option available for calculating 
$I_{ab}$ in a map containing only one cluster is to calculate
$I_{ab}$ assigning equal weight to each pixel in the map. However this 
procedure is prone to include nearby structures and objects 
that are not part of
the cluster, and can skew the resulting orientation of the principle axes.
A better scheme is to weight the the map with a function that dies off with
radius, such as a 2-D Gaussian distribution.  A circular Gaussian apodization
biases the computed ellipticity toward rounder values, and 
occasionally inaccurately identifies the principle axes.  
We therefore developed
a prescription to iterate the calculation of $I_{ab}$, replacing the 
initial circular Gaussian
window with progressively more elliptical 2-D Gaussians.

Since we assume that a real observation will rely on the BCG in 
the cluster to identify the axis direction, we choose to find the 
ellipticity and orientation of the innermost $400 \hspace{0.1 cm}h^{-1}\,{\rm kpc}$; we expect 
the gravitational environment at that scale to be the dominant 
contribution to the BCG orientation.  We first find the fractional
decrease in the average $\kappa$ at an annulus of radius $200 \hspace{0.1 cm}h^{-1}\,{\rm kpc}$ 
from the $\kappa$ at the
peak of the cluster.  We use this fraction to define the characteristic
size of each window in the iteration.  We begin with a circular 
2-D Gaussian window function, and use the eigenvalues and eigenvectors of 
the computed $I_{ab}$ to define the next elliptical window function in the 
iteration.  The iteration is continued until both the axis ratio and 
the slope of the axes have converged to a level of 0.1\%. To increase 
stability and speed, we have forced the change in ellipticity 
to be monotonic; if the ellipticity
in an iteration gets rounder, we elect to update only the slope.
To benchmark this prescription, we
have applied this technique to elliptical NFW profiles, and have
accurately recovered both the ellipticity and the slope.
For a small number of N-body clusters, a unique solution for slope and 
ellipticity could not be identified with this technique, usually 
due to merging objects that are not well characterized by an elliptical 
shape.  These objects are excluded from our analysis.

\begin{figure}
\begin{center}
\resizebox{5.5in}{!}{\includegraphics{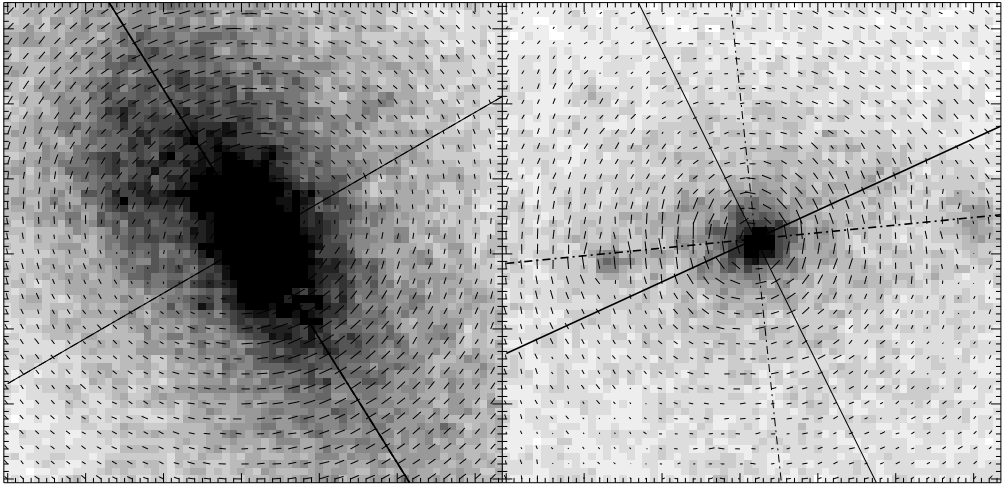}}
\end{center}
\caption{The $\kappa$ map, shear field, and principle axes for two isolated
N-body clusters. The boxes are $640\hspace{.1cm} \hspace{0.1 cm}h^{-1}\,{\rm kpc}$ 
on a side. The region affecting
the axis determination is somewhat greater than 
$400\hspace{.1cm} \hspace{0.1 cm}h^{-1}\,{\rm kpc}$ in diameter for more elliptical clusters. The
axes plotted are semi-major (bold) and semi-minor axes, where the dashed
lines are the initial eigenvectors of $I_{ab}$ and the solid lines are the 
result after the moment of inertia computation has been iterated
to a convergence threshold of \%0.1.}
\label{fig:smoothnlumpy}
\end{figure}

The convergence maps from two isolated clusters are
shown in figure \ref{fig:smoothnlumpy}.  The shear field 
is plotted over the top, as are the semi-major (bold) and semi-minor axes.  
The dashed lines are the axis directions after the first iteration and 
the solid lines are the converged determination of axis directions.  
Notice that the bold lines tend to point to regions of higher tangential
shear.
The cluster on the left has a moderately high ellipticity in the 
2-D plane and 
a relatively small amount of substructure in the region within 
$\sim 400 \hspace{0.1 cm}h^{-1}\,{\rm kpc}$ where the axes are being determined, thus the initial and 
converged solutions are virtually identical. 
In contrast the cluster on the 
right appears fairly round, and has substructure that affects 
the direction of the principle axes, requiring six times as many 
iterations to converge as the cluster on the left. 

\subsection{NFW profiles and isolated N-body clusters}\label{sec:analysis}
\begin{figure}
\begin{center}
\resizebox{4.5in}{!}{\includegraphics{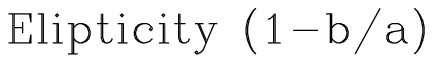}}
\end{center}
\caption{The computed $Q$ for the N-body clusters in our sample(crosses), as a 
function of the ellipticity $\varepsilon=1-b/a$.  We have used $W(\phi)
=\cos(2\phi)$.  The solid line shows the expected value of $Q$ for smooth 
NFW profile clusters. }
\label{fig:scatter}
\end{figure}

Using the iterator method to set the direction of
$\phi=0$ in the integral of equation
\ref{eqn:Qdef}, we can now compute the value of $Q$ for each cluster in 
our sample using $W(\phi)=\cos(2\phi)$.  For our mock observations of $Q$ 
we select an annulus between $800-1000 \hspace{0.1 cm}h^{-1}\,{\rm kpc}$; 
large enough 
to provide an adequate number of background galaxies while not so distant
as to lose the signal. We defer justification of this choice to section
\ref{sec:trade}.  Figure \ref{fig:scatter} shows $Q$
for each cluster in the sample, as a function of the computed 
ellipticity.  We define ellipticity as $\varepsilon=1-b/a$ where
$b$ and $a$ are the square roots of the smaller and larger 
eigenvalues of $I_{ab}$. 
Also plotted (solid line) is the value of $Q$ 
for smooth NFW profiles of the given ellipticity.  The NFW profiles used
are given by \cite{Bartelmann:1996hq}
\begin{eqnarray}\label{eqn:nfw}
\kappa(x)=\frac{2 \kappa_s f(x)}{x^2-1} 
\end{eqnarray}
where
\begin{eqnarray}
x=\frac{1}{r_s}\sqrt{x^2(1-\varepsilon)+
y^2/(1-\varepsilon)} 
\end{eqnarray}
and
\begin{eqnarray}
  f(x) = \begin{cases}
  1 - \frac{2}{\sqrt{x^2 - 1}}\hspace{.1cm}{\rm arctan}\sqrt{\frac{x-1}{x+1}} & (x > 1) \\
  1 - \frac{2}{\sqrt{1 - x^2}}\hspace{.1cm}{\rm arctanh}\sqrt{\frac{1-x}{1+x}} & (x < 1) \\
  1 & (x=1) 
\end{cases}
\end{eqnarray}
Here $\kappa_s=\rho_sr_s\Sigma_{cr}^{-1}$.  Both this parameter and the 
scale radius $r_s$ are set to their median values for our sample.
Values of $Q$ for both the NFW and N-body
clusters have been normalized with 
$\Sigma_{SIS}$, the 
convergence of a singular isothermal sphere at the radius of observation. 
Figure \ref{fig:scatter} demonstrates that $Q$ is correlated with 
ellipticity, but suffers from a large level of intrinsic scatter.  This 
scatter is initially 
somewhat surprising in light of the fact that the clusters are
being studied in isolation, with no projection effects or background 
noise added into the measurement.  The origins of the scatter are threefold;
the measurement of $Q$ is extremely sensitive to the existence of 
substructure in the observed annulus, and 
is moderately sensitive to how accurately 
the inner $\sim 400 \hspace{0.1 cm}h^{-1}\,{\rm kpc}$ is aligned with the cluster's 
orientation at $800-1000 \hspace{0.1 cm}h^{-1}\,{\rm kpc}$.  The determination 
of the ellipticity ($x$ axis)
is also affected by substructure in the inner $\sim 400 \hspace{0.1 cm}h^{-1}\,{\rm kpc}$ (see 
right panel of figure \ref{fig:smoothnlumpy}).

\begin{figure}
\begin{center}
\resizebox{5.5in}{!}{\includegraphics{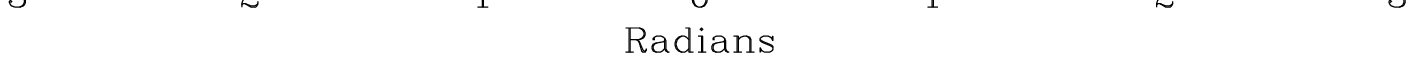}}
\resizebox{3.75in}{!}{\includegraphics{outlier.ps}}
\end{center}
\caption{{\bf TOP}  The asterisks show the azimuthal profile of
$(\gamma_T(\phi)-\bar{\gamma_T})/\bar{\gamma_T }$
for an isolated N-body cluster.  This profile
corresponds to the smoother cluster presented in figure \ref{fig:smoothnlumpy}. 
The heavy line depicts the azimuthal profile of an NFW cluster with the 
same ellipticity.  The wavy line show the profile of the object below, 
illustrating the impact of substructure.  The labels
A-F refer to features in the azimuthal profile, whose angular positions
are marked in the figure below.
{\bf BOTTOM}  This is one of the clusters in our 
sample that exhibits a high level of
substructure in the annulus being observed.  The circles have radii $200,
400, 800$ and $1000 \hspace{0.1 cm}h^{-1}\,{\rm kpc}$.  
The ellipse depicts the axis ratio identified via the 
moment of inertia in the core.}
\label{fig:pizza}
\end{figure}

The estimator $Q$ is designed to be sensitive to the azimuthal variation
of the tangential shear distortion.  This variation can be seen by eye in
the right panel of figure \ref{fig:smoothnlumpy}, wherein the 
elongation of the shear field vectors is more
pronounced near the pointy ends of the cluster.  In figure \ref{fig:pizza},
we plot the azimuthal variation of the left cluster in 
figure \ref{fig:smoothnlumpy}
(asterisks).  This cluster has an ellipticity of $\varepsilon=0.47$,
slightly more elliptical than the mean of our sample
($\varepsilon_{\rm mean}=0.414$) and has very little
substructure falling into the annulus $800-1000 \hspace{0.1 cm}h^{-1}\,{\rm kpc}$.
For comparison, we plot the azimuthal profile of a smooth NFW cluster 
with the same axis ratio (heavy line).
To illustrate the effect of substructure in the 
observational annulus, we also plot the azimuthal signal of the cluster shown 
in the bottom panel of figure \ref{fig:pizza} (wiggly line). 

Studying figure \ref{fig:pizza} yields some insight into the impact
of substructure on the observation.  In the bottom panel, 
the 4 circles are at $200$, $400$,
$800$ and $1000 \hspace{0.1 cm}h^{-1}\,{\rm kpc}$; the inner two mark the region most impacting the 
alignment of the BCG and our determination of the ellipticity, and the outer 
two mark the annulus being observed.   Some of the features of this cluster's
azimuthal profile are marked with letters. 
This object differs from the left hand
cluster of figure \ref{fig:smoothnlumpy} in that it is somewhat more elliptical,
thus the amplitude of the tangential shear is larger at point A.  Because 
there is very little substructure in the inner $400 \hspace{0.1 cm}h^{-1}\,{\rm kpc}$, our determination 
of the principle axes is accurate and the profile is not shifted to the
right or left in the top panel.
There is substructure at points B, D, E, and F.  In the vicinity of these
blobs, the shear signal encircles the feature, instead of being tangential
to the main part of the cluster.  This results in a signal enhancement near 
the center of the blob, and a suppression to its left and right, where the 
shear is essentially radial with respect to the cluster center.  Substructure
causes an effect even when it's outside the annulus, as in D, but the impact 
is smaller.  At point C, 
the suppression from B and D conspire to create a huge feature in the
profile.  The substructure at B and E will increase the net $Q$ because
$W(\phi)$ is positive in those places.  The substructure at points D and F 
will decrease the net $Q$.  
Although it is not shown in this example, substructure in the inner regions 
can cause the BCG to be misaligned with the bulk of the cluster.  This would
cause a horizontal shift in the profile plotted in figure \ref{fig:pizza}, 
and would have the overall effect of decreasing $Q$ in an observed measurement.

\subsection{The optimal weight function}\label{sec:optimal}

In noting the departure
of the NFW azimuthal profile from the 
quadrupolar variation $\cos(2\phi)$ (figure \ref{fig:pizza}), 
it is interesting to 
determine whether using a modified shape of the weight function $W(\phi)$
results in a higher quality signal. To this end, we have identified an
analytic form that describes 
the shape of the azimuthal profile for the smooth NFW clusters. 

\begin{equation}\label{eqn:nfwfit}
F(\alpha,\phi)=\frac{1}{\alpha}\left(e^{\alpha\cos
(2\phi)}-1\right)-P(\alpha)
\end{equation}
The parameter $\alpha$ monotonically increases with
ellipticity.  $P(\alpha)$ is an offset that 
vanishes as $\alpha \rightarrow 0$, so that $F(\alpha) \rightarrow 
\cos(2\phi)$ in that limit.
The precise dependence 
of $\alpha$ and $P(\alpha)$ on ellipticity depends
on the radius and width of the annulus being observed.  

In practice, to test a 
modified $W(\phi)$, we fit both $\alpha$ and the offset for a single 
NFW ellipticity, with axis ratio equal to the mean of our cluster sample,
$b/a=0.586$.  We used this modified weight 
function to compute an alternate estimator, $Q'$, and compared it to 
$Q$ with mixed results.  On the one hand, for many of the 
isolated N-body clusters 
the strength of the signal is significantly increased, especially 
for the most elliptical objects. However the intrinsic scatter seen in 
figure \ref{fig:scatter} is not improved, and the
performance of $Q'$ in the presence of contaminants (see section 
\ref{sec:contam}) such as instrument 
noise, projection effects, and misalignment of the BCG is substantially 
worse than when we use $W(\phi)=\cos(2\phi)$.  The peakier
profile makes $Q'$ much more sensitive to misalignment of the 
semi-major axis.  Also, since the modified $W(\phi)$ 
does not integrate to zero, 
several of the less elliptical clusters with $b/a>0.586$ wind up with a negative
value of $Q'$, which  results in a lower mean signal for the sample, and
is also difficult to interpret for an individual cluster.  

\subsection{Noise, projection, and axis misalignment}\label{sec:contam}
\begin{figure}
\begin{center}
\rotatebox{90}{\resizebox{3.75in}{!}{\includegraphics{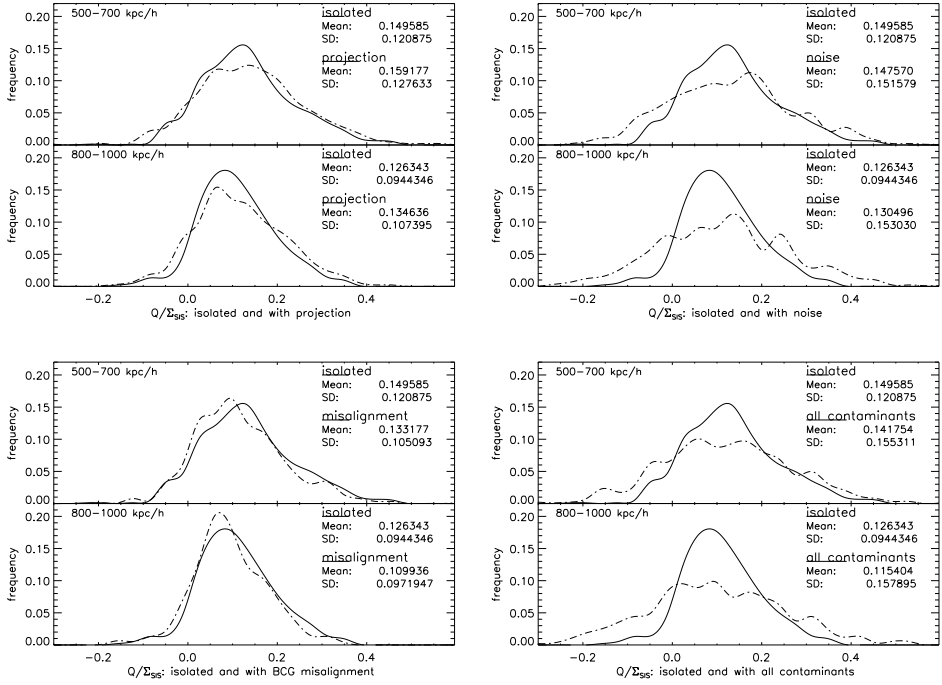}}}
\end{center}
\caption{The histograms demonstrate the
 impact of contaminants on the expected distribution are 
shown for two observational annuli.  The effects of projection (top left)
noise (top right) BCG misalignment (bottom left) are shown separately, and 
in combination (bottom right). Solid lines mark the distribution without 
any contamination.}
\label{fig:hist}
\end{figure}

In practice, a measurement of the quantity $Q$ for real clusters in the 
universe will be made in the presence of a variety of contaminants.  
There are a finite number of background galaxies that can be used to measure
the shear distortion.  They will not be evenly distributed, and 
the number found will depend on the depth of the observation.  Galaxy 
redshifts will be impacted if they fall close to
central BCG or are eclipsed by foreground galaxies. 
The instrument will also
introduce noise in the shear measurement.
As is well known, weak lensing measurements suffer from projection
effects from objects that lie along the line of sight \cite{dePutter:2004xp}, 
which in this
context is important as these projections
have a similar impact on the signal as substructure. Finally, while we have 
assumed that the BCG will be aligned with the underlying dark matter 
distribution, in reality its orientation will depend substantially on its 
merger history, and it has been suggested that under some circumstances, 
its orientation does not trace that of the larger halo very 
well \cite{Gonzalez:2004ah}.
It is important to include these contaminants when predicting an expected 
distribution of $Q$ observations.  In figure \ref{fig:hist}, we show how
projection, noise and misalignment individually impact the expected 
distribution, and also summarize the result when all three are present. 
These results are for a sample of $\sim$900 projections 
through 30 N-body clusters.
The histograms are in $26$ bins over the domain plotted.  The distribution
for isolated clusters contains values of $Q<0$ because of the influence of
substructure (see section \ref{sec:analysis}).

To model projection, we have added the convergence map of the isolated 
cluster to a line of sight projected $\kappa$ map of the same angular 
size (see section \ref{sec:sims}).  Because this method neglects the fact 
that galaxy clusters often exist in regions of intersecting large scale 
structure, we expect that we have slightly underestimated the impact of
projection effects, but believe our analysis provides a firm lower bound 
on the extent of the effect.  Objects in the line of sight increase the 
mean of the distribution by around 7\%, and increase the scatter by 
about 14\% in the larger bin.  The scatter is affected much less for 
smaller observational radii,
because there is less area in the annulus. 

To model the effects of instrument noise and variance in the number of
available background galaxies, we have chosen to add a conservative 
level of white noise to the observation.  We
postulate that there will be fluctuation per component of the
reduced shear of $(\delta\gamma)_{\rm rms}=0.2$ 
(where $\gamma_{\rm red} = \gamma/(1-\kappa)$), and that there
will be a mean number of background galaxies $\bar{n}_{\rm gal}\sim 100$ 
per arcminute with
which to make a shear measurement.   Therefore the noise that we add to
each component to get the observed shear is given by
\begin{eqnarray}
\gamma_{\rm obs}=\gamma \pm \sqrt{\frac{(\delta\gamma)^2}{N_{\rm gal}}}
\qquad {\rm with}\ N_{\rm gal}=\bar{n}_{\rm gal}\theta_{\rm pix}^2
\end{eqnarray}
where we ignore intrinsic alignments of the background galaxies, which
will be a second order effect.
The effect of this noise on the distribution of measured $Q$ values can
be seen in the upper right panel of figure \ref{fig:hist}.
The impact is that in the $800-1000 \hspace{.1cm} h^{-1}\,{\rm kpc}$ bin the 
standard deviation is increased by 62\% from the inherent scatter. 
Comparing this to the effects of projection (upper left)
and misalignment (lower left), it is clear that this noise is the dominant
contribution to the signal contamination, even though achieving a level
of $\bar{n}_{\rm gal}\sim 100$ per arcminute 
will require deep observations from a space based platform.  

Perhaps the weakest element in this approach to measuring cluster shape is 
the reliance on the alignment of the central BCG with the underlying dark 
matter distribution in the cluster.  We have studied the impact of
misalignment on the $Q$ values of NFW profile clusters and find that for 
an NFW profile of mean ellipticity $\varepsilon=0.414$, a misalignment of 
$10^\circ$ leads to a $5$\% decrease in the measured $Q$, while a 
$20^\circ$ leads to a $30$\% decrease. 
In analyzing the N-body simulations, we have tried to be realistic in 
our determination of the principle axes, in that we have used the inner $\sim
400 \hspace{0.1 cm}h^{-1}\,{\rm kpc}$ to calculate the expected BCG alignment angle. 
This means the alignment 
will be sensitive to the gravitational 
effects of in-falling substructure near the core.  However, since the 
details of BCG alignment are not known, we have 
introduced a random scatter in BCG orientation from the 
eigenvectors of the moment of inertia tensor.  
The impact of a $15^\circ$ rms scatter in the BCG orientation on the
expected distribution of the observable can be seen in lower left
panel of figure \ref{fig:hist}.  

As predicted, introducing misalignment systematically lowers the mean value
of $Q$ because the weight function in the integral of equation \ref{eqn:Qdef}
will have  a phase lag with respect to the 
azimuthal variation of the shear.  In the $800-1000 \hspace{.1cm}
h^{-1}\,{\rm kpc}$ bin the signal is decreased by around 13\%.  It is 
interesting to note that the impact on
the spread of the distribution is small.  This estimator using
$W(\phi)=\cos(2\phi)$ is substantially more robust in the presence of BCG
misalignment than one that uses a 
peakier weight function of the type in equation 
\ref{eqn:nfwfit} (denoted $Q'$ above), because $\cos(2\phi)$ 
provides more overlap with the true signal in the presence of a phase lag.
Compare the bold and dashed profiles in figure \ref{fig:pizza} to see this. 

The bottom right panel of figure \ref{fig:hist} shows the expected 
distribution of measured $Q$ values in the presence of all the contaminants. 
While this distribution is quite broad, it is still a statistically 
significant signature of the ellipticity of galaxy clusters.  The mean
ellipticity in our sample of N-body clusters 
is $\varepsilon_{\rm mean}=0.414$, however, connecting the measured value of $Q$ to 
an effective ellipticity must be done with caution, because substructure
causes a large intrinsic scatter in the $Q - \varepsilon$ 
relationship.
Of the contaminants, the most significant is noise in the shear measurement.   
The standard deviation of $Q$ values is increased by around 67\% from
the intrinsic level of scatter, and
the mean of the distribution is decreased by 9\% in the  
$800-1000 \hspace{0.1 cm}h^{-1}\,{\rm kpc}$ bin.

\subsection{Mock observations and noise tradeoffs}\label{sec:trade}
\begin{figure}
\begin{center}
\resizebox{3.75in}{!}{\includegraphics{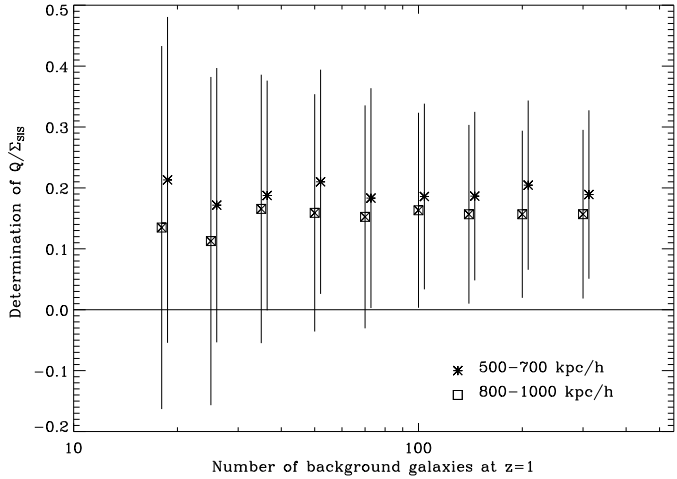}}
\end{center}
\caption{Mean and scatter of the measured estimator as a function of 
the number of background galaxies in the observation.  Results are 
shown for annuli at $500-700 \hspace{0.1 cm}h^{-1}\,{\rm kpc}$ 
(asterisks) and $800-1000 \hspace{0.1 cm}h^{-1}\,{\rm kpc}$
(squares).  Error bars depict 1$\sigma$ values.
Asterisks have been shifted slightly right of the true abscissa
for visual clarity. }
\label{fig:qvsngal}
\end{figure}

Figure \ref{fig:qvsngal} shows the results of a mock observation of 
the estimator as a function of the number of background galaxies.  The mean 
values of $Q$ are plotted along with $1\sigma$ error bars indicating the 
significance of the detection of asphericity on a single measurement.
Results are shown for observations performed in two 
separate annuli, $500-700 \hspace{0.1 cm}h^{-1}\,{\rm kpc}$ and $800-1000 \hspace{0.1 cm}h^{-1}\,{\rm kpc}$. The abscissa are 
the same values for each, but we have shifted the asterisks for visual clarity. 
The plot demonstrates that there are diminishing returns for increasing 
the depth of the observation to exceed $\sim 100$ background galaxies per
square arcminute.
Past this level, the errors are increasingly dominated by projection, 
BCG misalignment, and intrinsic scatter in the value of $Q$ due to 
cluster substructure.  This scatter cannot
be improved by changing the experiment design unless a better method 
of detecting the orientation of the dark matter halo is discovered. 
Many ground based experiments that propose to measure the shear distortion
will not be able to reach a depth of $100\,{\rm arcmin}^{-2}$.
Figure \ref{fig:qvsngal} implies that in order for these experiments to 
rule out a spherical dark matter halo using weak lensing shear, 
they will have to observe closer to 
the center of the cluster.  In practice, this will be difficult to achieve 
because there will be more confusion in the shear measurement due to 
foreground galaxies in the cluster. 

\begin{figure}
\begin{center}
\resizebox{3.75in}{!}{\includegraphics{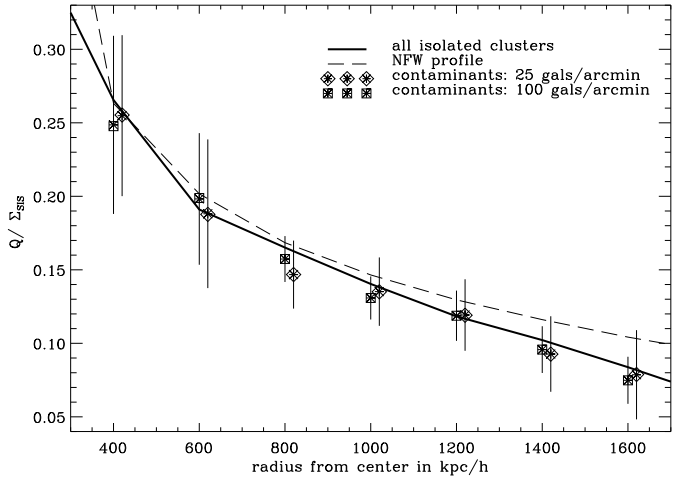}}
\end{center}
\caption{The radial profile of the estimator $Q$ for isolated N-body 
clusters (solid), an NFW profile of average ellipticity (dashed), and 
mock observations of 900 clusters, with contamination from noise, 
projection effects, and misalignment of the BCG. Noise levels at
100 background galaxies (squares), and 25 background galaxies (diamonds,
shifted right).  Error bars 
are bootstrapped $3\sigma$ values, and the mock observations are made 
in non-overlapping annular bins of width $200 \hspace{0.1 cm}h^{-1}\,{\rm kpc}$.}
\label{fig:boots}
\end{figure}

Figure \ref{fig:boots} shows the radial profile of $Q/\Sigma_{SIS}$ for 
a sample of around 900 clusters.  The solid line shows the radial profile of the 
the isolated clusters.  This result is averaged over the sample; the individual
radial profiles exhibit large deviations from this mean. 
The dashed line indicates the radial profile of a smooth
NFW cluster, with the mean ellipticity, and median values for 
the scale radius and $\kappa_s$ (see equation \ref{eqn:nfw}). 
The squares indicate the results for
a mock observation of about 900 clusters, at a noise level of 100 background 
galaxies per arcminute, and the diamonds show the result for 25 
background galaxies per arcminute.  The error bars are bootstrapped from 700 
synthetic data sets, each derived from the original data 
by discarding $\sim 330$ clusters with replacement.  Error bars are plotted
at $3\sigma$.  The mean values of the mock observation are lower than those
of the isolated clusters because random misalignments systematically lower
the signal.  

Figure \ref{fig:boots} suggests that an observation of 900 clusters at 
25 background galaxies per arcminute can successfully 
observe cluster asphericity.
It also suggests that with an observation of 900 clusters
at a noise level of 100 back ground galaxies per arcminute, it may be 
possible to distinguish between the predictions of N-body CDM simulations,
and those of hydrodynamic simulation in \cite{Kazantzidis:2004vu}, which 
predict that the ellipticity of dark matter halos is reduced by 20\% with
the inclusion of baryonic physics.  We draw this conclusion loosely, 
based on the observation that the $Q$ of an NFW cluster of
smaller ellipticity $\varepsilon=0.331$ is reduced by $24-25$\% from
the values shown in figure \ref{fig:boots}. 
However, we caution the 
reader that these conclusions are somewhat dependent on our assumptions about
BCG misalignment.  BCG misalignment systematically reduces the measured 
value of $Q$, and is therefore degenerate in its effect with rounder 
clusters.  Another note of caution, we have seen that substructure in 
the cluster profoundly affects the measured value of $Q$.  The impact 
of baryonic physics on substructure is that cooling allows
deeper potential wells to form, which may increase the influence 
of substructure on the 
measurement.  We suggest that changes in the character of substructure
may be as important in this type of observation as the changes
in the overall shape of the halo.  Quantifying this conjecture, however, 
is beyond the scope of this paper. 

\section{Conclusions and future work}

This work has been a study in identifying an observable estimator of galaxy
cluster shape.  The estimator, $Q$, measures the azimuthal variation of 
tangential weak lensing shear. Using N-body simulations, we have computed
the expected mean value and distribution of this quantity.  
We show that while $Q$ is definitely correlated with ellipticity, there
exists a great deal of intrinsic scatter in the relation. 
 
The dominant contribution to the intrinsic scatter is the presence of
substructure in the cluster.
We have carefully examined the impact of substructure on this estimator
of cluster shape.  We conclude that substructure at small radii can cause
misalignment of the orientation of the innermost region with that of the
larger dark matter distribution of the cluster.  If this misalignment
is reflected in the orientation of the BCG, substructure near the core is
expected to decrease the measured value of $Q$, however, the factors that
determine the BCG alignment are not well understood.
Substructure in the observed annulus can either increase or decrease the
measured value of $Q$, depending on its azimuthal position with respect
to the principle axes.
Its effect is to increase the intrinsic level of scatter.
It is possible that finding substructure in advance, using a method
(e.g.~flexion) that is sensitive to small-scale variations in the shear,
and then removing it would reduce this source of noise.  How well this
could be performed on low-S/N data, such as we are likely to have on a
per cluster basis, is unclear.

We explored the possibility that the estimator could be improved by
matching the azimuthal weight function $W(\phi)$ to the shape of the
variation for an NFW cluster of average ellipticity.  We find that an
estimator of that construction often results in a larger signal for
many clusters, but is much less robust in the presence of
contaminants, and more difficult to interpret than $\cos(2\phi)$.

There are a few aspects that we have not explored in this work.
We have not varied the width of the observational annulus, though
our analysis suggests that a wider annulus may be viable
(see figure \ref{fig:boots}).  We believe that there is useful
information in the radial dependence of $Q$, and that increased S/N can
always be obtained by averaging the noisier results post facto, which we
take as arguments in favor of the narrower annuli we have chosen.
We have neglected intrinsic alignments of background galaxies in our treatment
of noise, and we have not accounted separately for
uncertainties in the redshifts of lenses and sources.
We have not sampled large scale structure accurately when including the
effects of projection, which has probably caused us to somewhat underestimate
the impact.   We have relied heavily on the hypothesis that
the BCG is somewhat aligned with the dark matter in the galaxy cluster.
Finally, we have drawn some loose conclusions about the efficacy of this
approach in differentiating the predictions of CDM and the hydrodynamic
simulations of \cite{Kazantzidis:2004vu}, but we note that these are likely
to be altered by the influence of baryons on substructure.  We hope that
future work will address these issues.

We have nonetheless made realistic predictions by
including the effects of noise, light cone projections, and misalignment
of the central galaxy on the distribution of $Q$.  We conclude that 
the contaminants do not entirely erase the signal, and that a 
statistical measurement of cluster ellipticity is tractable.
We show that even for a reasonably deep observation with 
100 background galaxies
per square arcminute pixel, observation noise is still the dominant source of 
uncertainty over the intrinsic level of scatter.  We have quantified the
relation between the number of background galaxies and the significance
of the detection.  We have examined the radial profile of $Q$, and 
quantified the statistical significance of observations with 25 and 
100 background galaxies per arcminute.   We conclude that both 
resolutions should be able to detect cluster asphericity given a 
sample size of 900 clusters like that analyzed here.  
The higher resolution measurements 
that can be made with a next generation space-based platform such as SNAP
may be able to distinguish between a mean ellipticity of $\varepsilon=0.414$
and one that is 20\% smaller, enabling us to measure the claimed effects of
baryonic physics on cluster ellipticities.

\bigskip

Many thanks to Neal Dalal, Paul Bode, 
Chris Vale,Masahiro Takada, and Daniel Podolsky for useful discussions 
about this work.  This research was supported in part by grant 
NSF-AST-0205935, other NSF grants, and NASA.  JFH
is supported by NASA through Hubble Fellowship grant \#
01172.01-A, awarded by the Space Telescope
Science Institute, which is operated by the Association of
Universities for Research in Astronomy, Inc., for NASA, under contract
NAS 5-26555. The simulations were performed on facilities provided
by the National Center for Supercomputing Applications (NCSA).
\bigskip
\bigskip


\end{document}